\documentclass{JHEP3}

\preprint{LU-TP 06-29\\
  hep-ph/0608210\\
}

\usepackage{color}
\let\normalcolor\relax
\usepackage{cite}
\usepackage{epsfig}
\usepackage{xspace}
\usepackage{graphics}
\usepackage[latin1]{inputenc}

\def\hide#1{}

\newcommand{\pythia}{P\scalebox{0.8}{YTHIA}\xspace}

\newcommand{\ET}{\ensuremath{E_{\perp}}}

\newcommand{\pTi}[1]{\ensuremath{p_{\perp #1}}}

\newcommand{\ETI}[1]{\ensuremath{E_{\perp \mathrm{#1}}}}

\def\mrm#1{\mathrm{#1}}

\def\sub#1{\ensuremath{_{\mrm{#1}}}}

\def\f2d3{\ensuremath{F_2^{\mrm{D}3}}}

\def\done#1{}

\newcommand{\eqref}[1]{eq.~(\ref{#1})\xspace}
\newcommand{\eqsref}[1]{eqs.~(\ref{#1})\xspace}

\newcounter{aenumct}

{\end{list}}

\newcounter{enumct}

 \renewcommand\email[1]{{\scriptsize\tt\href{mailto:#1}{#1}}}

\newcommand{\mbar}{\ensuremath{\overline{m}}}
\newcommand{\lbar}{\ensuremath{\overline{l}}}
\newcommand{\bperp}{\ensuremath{\bar{b}_{\perp}}}

\newcommand{\kperp}{\ensuremath{\bar{k}_{\perp}}}

\newcommand{\Aborn}{\ensuremath{A_{\rm{Born}}}}
\newcommand{\Aeik}{\ensuremath{A_{\rm{eik}}}}

\def\mrm#1{\mathrm{#1}}
\def\sub#1{\ensuremath{_{\mrm{#1}}}}

\def\mpl{\ensuremath{M\sub{P}}}
\def\mpln#1{\ensuremath{M\sub{P #1}}}
\def\rsch{\ensuremath{r\sub{Sch}}}

\def\mmin{\ensuremath{M\sub{min}}}
\def\mminn#1{\ensuremath{M\sub{min #1}}}
\def\Meff{\ensuremath{M\sub{eff}}}
\def\Mjj{\ensuremath{M_{jj}}}
\def\MBH{\ensuremath{M\sub{BH}}}
\newcommand{\figref}[1]{fig.~\ref{#1}\xspace}
\newcommand{\figrefi}[1]{fig.~\ref{#1}}

\def\Ms{\ensuremath{M_s}}
\def\MD{\ensuremath{M_D}}
\newcommand{\charybdis}{C\scalebox{0.8}{HARYBDIS}\xspace}

\skip\footins = 1\bigskipamount plus 2pt minus 4pt                              

\title{\boldmath Classical and Non-Classical ADD-phenomenology with
  high-\ET\ jet observables at collider experiments}

\author{Leif Lönnblad and Malin Sjödahl\\
  Dept.~of Theoretical Physics,
  S\"olvegatan 14A, S-223 62  Lund, Sweden\\
  E-mail: \email{Leif.Lonnblad@thep.lu.se}
    and \email{Malin.Sjodahl@thep.lu.se}}
  
  \abstract{We use the results from a recent investigation of hard
    parton--parton gravitational scattering in the ADD scenario 
    to make semi-quantitative predictions for a
    few standard high-\ET\ jet observables at the LHC. By
    implementing these gravitational scattering results in the \pythia event
    generator and combining it with the \charybdis generator for black
    holes, we investigate the effects of large extra dimensions and
    find that, depending on the width of the brane, the relative
    importance of gravitational scattering and black hole production
    may change significantly. For the cases where gravitational
    scatterings are important we discuss how to distinguish gravitational
    scattering from standard QCD partonic scatterings. In particular
    we point out that the universal colorlessness of elastic 
    gravitational scattering implies fewer particles
    between the hard jets, and that this can be used in order to 
    distinguish an increased jet activity induced by gravitational scattering
    from an increased jet activity induced by eg.\ super-symmetric extensions
    where the interaction is colorful.}

\begin{document}
 
\sloppy
 
\section{Introduction}
\label{p6:sec:intro}

The most exotic, and by far most discussed collider signal of large
extra dimensions in the ADD scenario
\cite{Arkani-Hamed:1998rs,Arkani-Hamed:1998nn,Antoniadis:1998ig} is the
copiously produced extra dimensional black holes 
\cite{Emparan:2000rs, Dimopoulos:2001hw,Giddings:2001bu,Kanti:2004nr}. 
While these are
expected to come with large cross sections and characteristic signals
at the LHC for a ``natural'' Planck scale of around 1~TeV, both cross
section and signals suffer severely from uncertainties associated with
quantum gravity. This provides a wonderful chance to probe quantum
gravity, but from the point of view of verifying the scenario it is
not ideal.  It is therfore worth looking for processes which involve
fewer uncertainties than decaying black holes.

Other important ADD processes involve Kaluza--Klein modes, 
either the production of real ones, or the exchange of virtual 
ones in gravitational scattering of hard partons. 
The signal for the former involves a large
missing transverse momentum and will be difficult to distinguish from
eg.\ the production of stable super-symmetric particles in some SUSY
extensions of the standard model. The later will show up as an
increase of the jet cross sections at high energies and may, if this increase is is small, be difficult to distinguish from other 
beyond-the-standard-model effects.

In a previous paper \cite{Lonnblad:2005ah} we investigated an
alternative signal for the ADD scenario, namely the disappearance of
the high-\ET\ jet cross section due to the formation of black holes.
However, in that paper we neglected the contribution from hard
gravitational scattering.

Lately a coherent picture of gravitational scattering in the ADD model,
at both low and high energies, was presented in \cite{Sjodahl:2006gb}.
In this paper we investigate the phenomenological consequences. Again,
we will concentrate on standard jet observables to see how they are
affected by the existence of large extra dimensions, using different
choices of the model parameters. We will try to give a complete
semi-quantitative description of the observables ranging from the
region of perturbative gravitational scattering in the low-energy end
to the domain of classical (non-quantum gravitational) black holes for
energies above the Planck scale.

While the LHC should easily discover large extra dimensions for the most
natural choices of Planck masses and number of extra dimensions, we
find situations where no black holes are formed and the only gravitational 
scattering signal could be a slight increase of
the \ET\ and di-jet cross section at high energies. 
We therefore discuss the possibility of distinguish such
scatterings from standard QCD events by studying the different color
topologies involved. We also suggest that such a procedure could be
used at the Tevatron to see if an increase of the high-\ET\ jet cross
section there could be the result of the onset of subplanckian
gravitational scattering.
 
This paper is organized as follows. After a brief introduction to the
ADD scenario in section \ref{p6:sec:add}, we summarize in section
\ref{p6:sec:effgrav} the description of gravitational scatterings
developed in \cite{Sjodahl:2006gb}. We then go on to discuss the
production and decay of black holes in section \ref{p6:sec:BH} and, in
section \ref{p6:sec:results}, we present our results before discussing
our conclusions in section \ref{p6:sec:conclusions}.

\section{Basics of ADD}
\label{p6:sec:add}

The so called ADD scenario, invented in 1998 by Arkani-Hamed, Dvali
and Dimopoulos \cite{Arkani-Hamed:1998rs,Arkani-Hamed:1998nn}, aims at
explaining the hierarchy problem, i.e.  why the observed Planck scale
at $10^{19}$~GeV is so large compared to the masses of the standard
model particles. This is done by introducing a number, $n$, of extra
dimensions in which only gravity is allowed to propagate.

In order to explain why these dimensions have not yet been observed, it
is assumed that they are compactified with some (common)
compactification radius $R$\footnote{We use R to denote the
  compactification radius rather than the compactification
  circumference (see the appendix for a discussion on conventions). }
and that no gauge fields are allowed to
propagate in the extra dimensions.  Gravity, on the other hand is, and
this renders the form of Newton's law at distances, $r$, much smaller than
the compactification radius
\begin{eqnarray}
  \label{p6:eq:V}
  \frac{V(r)}{m_1m_2}
  &=& -\frac{S_n \Gamma(n)}{\mpl^{n+2} (2 \pi)^n} \frac{1}{r^{n+1}}.
\end{eqnarray}
Here $\mpl$ is the fundamental Planck scale, $S_n=2\pi^{n/2}/\Gamma(n/2)$ 
is the surface of a unit sphere in $n$ dimensions 
and $\Gamma(n)$ is the Euler Gamma function.
At distances large compared to the compactification radius we must recover
the normal 3+1-dimensional form of Newton's law.  
\begin{eqnarray}
  \label{p6:eq:Vlarger}
 \frac{V(r)}{m_1m_2}
  &=& 
 -G_{N(4)} \frac{1}{r}.
\end{eqnarray}
Expressing Newton's constant in terms of the observed (3+1)-dimensional 
Planck scale, $G_{N(4)} \sim 1/\mpln{4}^2$, then
gives the relation $\mpln{4}^{2} \sim \mpl^{n+2}R^n$ between the 
fundamental Planck scale $\mpl$ and the observed 4-dimensional Planck
scale, which explains how the fundamental Planck scale could be (almost)
of the same order as the weak scale, but the observed  effective
Planck scale, $\mpln{4}$, many orders of magnitude larger.

However, this also implies that gravity should be very strong at small 
distances which opens up for the possibility of observing gravitational
scattering and black holes at collider experiments. 

\section{Gravitational scattering in ADD scenario}
\label{p6:sec:effgrav}

Although the field theory of gravity is ultimately divergent also in more
than 4 dimensions, an effective low-energy theory can be constructed by
a perturbative treatment of the metric in the limit where the metric
perturbation is small. A Lagrangian can be derived and Feynman
diagrams can be constructed from it. This is done in
\cite{Han:1998sg,Giudice:1998ck}. Since the extra dimensions are
compactified, momentum occurs in each direction as multiples of some
ground frequency, ie.\ as Kaluza--Klein modes.

In a gravitational event an outgoing Kaluza--Klein (KK) mode will have
some (quantized) momentum in the extra dimensions, which enters in the
(3+1)-dimensional Lagrangian as a mass term.  But the KK modes can also occur 
as intermediate states in which case they have to be properly summed or,
taking the continuum limit, integrated over. This gives rise to
the integral
\begin{equation}
  \label{p6:eq:mint}
  \sum_{\mbar_{\lbar}}{\frac{1}{-m_{\lbar}^2+k^2}} 
  \approx S_n R^n \int{\frac{m^{n-1}}{-m^2+k^2}}dm.
\end{equation}
Here $\lbar$ enumerates the allowed momenta, $m_{\lbar}$, in the extra
dimensions, $m$ is the absolute value of $m_{\lbar}$, and $k^2$ is the
momentum squared of the $3+1$-dimensional part of the propagator.
Note that this sum over KK states does imply momentum non-conservation
for momenta transverse to the brane where the standard-model fields 
live, but this is not a complete surprise since translational invariance 
is broken in the bulk by the presence of the brane.

\subsection{Dealing with divergences}
\label{p6:sec:div}

What is worrying though, is that the field theory seems to contain
divergences already at the tree level. However, the divergences
dissappear when imposing the requirement that the standard model
particles live on a brane, either directly by assuming a narrow
distribution of the standard model fields into the extra dimensions
\cite{Sjodahl:2006vq,Sjodahl:2006gb}, or by a introducing a ``brane
tension'' \cite{Bando:1999di,Kugo:1999mf}.  Both these methods gives
physical effective cut-offs for the momentum (mass) of the KK modes.
For example, a Gaussian extension $e^{-m^2/(2 M_s^2)}$ of the
standard model field densities into the bulk gives an ``effective propagator''
\cite{Sjodahl:2006gb}
\begin{equation}
  \label{p6:eq:SMprop}
  D(k^2)=R^n S_n\int \frac{dm \,m^{n-1}}{k^2-m^2} e^{-m^2/\Ms^2}
\end{equation}
for the exchange of KK modes with four-momentum exchange $k^2$.  (This
object, $D(k^2)$, is here sloppily called a propagator, despite the fact
that the multiplicative Lorentz structure is not taken into account.)

For momentum exchange small compared to $\Ms$, the standard model 
momentum $k$ in the propagator is irrelevant (for most $m$ in the integral), 
such that s-, t-, and u-channels are equally efficient and the scattering 
is fairly isotropic.

For $\sqrt{k^2} \gg \Ms$ on the other hand, the interaction is
dominated by forward scattering via the t-channel, and an all-order
eikonal calculation is necessary to ensure unitarity
\cite{Giudice:2001ce,Sjodahl:2006gb,Nussinov:1998jt}. 
The stage is therefore set by
three energy scales, the fundamental Planck mass, $\mpl$, the inverse
brane width (brane tension), $\Ms$, and the rest mass of the partonic scattering, $\sqrt{s}$,
and the phenomenology depend on their relative
magnitude.  It is illuminating to fix one of these scales an
study the different kinematical regions in the plan spanned by the
other two. This is done in \figref{p6:fig:sMs}where the
$(\sqrt{s},\Ms)$-plane is plotted for $\mpl$ fixed to 1 TeV.

Below we will successively describe the contribution from the t-, u-,
and s-channels and the various regions in \figref{p6:fig:sMs}.

\FIGURE[t]{%
      \epsfig{file=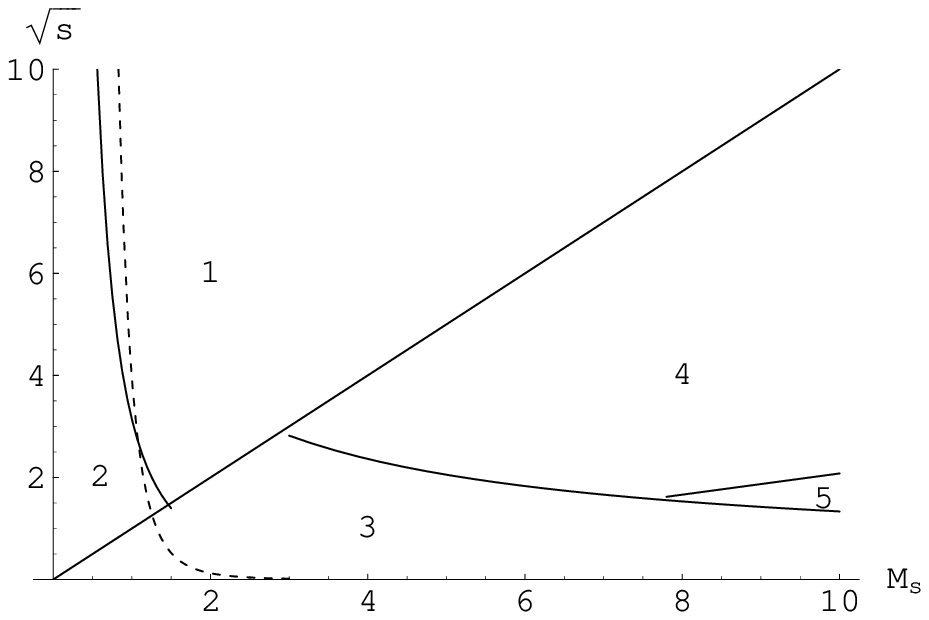,width=10cm}
\caption{\label{p6:fig:sMs} The $(\sqrt{s},\Ms)$-plane for $n=4$ and $\mpl=1$. 
The straight line separating region 1 and 4 is $\sqrt{s}=\Ms$ 
while straight line separating region 4 and 5 is the line where the real 
and imaginary parts in \eqref{p6:eq:1-loop} have equal magnitude.
The power-like solid curve separating region 1 and 2 is $\sqrt{s_{c}}$ from 
\eqref{p6:eq:scd} as a function of $\Ms$ and the line separating the regions 
4 and 5 from region 3 is the line where $|A_{\rm{Born}}X|=1$, 
see \eqref{p6:eq:1-loop}. 
In the regions 
1 and 2 $\sqrt{s}$ is larger than $\Ms$, and, at least for $\sqrt{s} \gg \Ms$, the 
eikonal approximation is correct. In region 1 the eikonal is, depending 
on $b$, either large compared to 1 or given by \eqref{p6:eq:chibigb}. 
In region 2 on the other hand 
the $b$-range where $|\chi|$ is small includes a region where it is
described by \eqref{p6:eq:chismallb}. In region 3 the correction 
corresponding to higher order loops is small, but in region 4 it is 
important and helps assuring unitarity. 
The dashed line indicates the minimal $\sqrt{s}$ (for a given $\Ms$) at which
the black hole radius \eqref{p6:eq:rSch} is larger than the brane width.
The plot visibly very similar for n=6.
} }

\subsection{t-channel}

As argued in the above section, we expect t-channel contributions to
dominate at energies high compared to $\Ms$. Unitarity constraints
does, however, imply that the Born approximation can not be valid for
sufficiently high energies. In fact, as is argued in
\cite{Giudice:2001ce,Landau:1987gn}, a completely new phenomenon 
occurs for scattering in more than $3$ spatial dimensions; 
namely the emergence of a length scale associated with the transition 
from the classical to the quantum domain.

Intuitively this can be understood by considering the ratios
\begin{equation}
  \label{p6:eq:theta}
  \frac{\Delta \theta}{\theta} \mbox{ and }
  \frac{\Delta b}{b} 
\end{equation}
where $\theta$ is the scattering angle and $b$ the impact parameter
in a scattering experiment. In the classical domain these ratios 
are both much smaller than $1$. Requiring the opposite, and approximating
\begin{equation}
  \label{p6:eq:theta2}
  \Delta \theta 
  \sim \frac{\Delta q}{M v}
  \sim \frac{\hbar}{M v \Delta b}
  \mbox{ and }
  \theta 
  \sim \frac{b}{Mv^2}\frac{dV(b)}{db}
\end{equation} 
for a non-relativistic particle with speed, $v$, mass, $M$, and
transverse momentum, $\Delta q$, moving in a potential, $V$, one
finds, for a Coulomb-like potential $V(b)=\alpha / b$,
the condition $\alpha < \hbar v$.  For coupling constants close to
$1$, this basically implies that the relativistic and quantum
mechanical regions coincide. For a more general potential of the form
$V(b)=\alpha / b^{n+1}$, assuming $n$ to be positive (this is what
Gausses law gives in 3+n spatial dimensions), the separation of the
classical and quantum domain depends on the impact parameter, such
that, the transition occurs at $b_c\sim [\alpha / \hbar v]^{(1/n)}$.
For gravitational coupling with $\alpha=G_{4+n}MM$, this 
corresponds to $b_c \sim [G_{4+n} M^2 /(v\hbar)]^{1/n}$
\cite{Giudice:2001ce}. Scattering in the potential \eqref{p6:eq:V}
is therefore expected to be mainly classical
only if the impact parameter $b$ is smaller than $b_c$.

\FIGURE[t]{%
      \epsfig{file=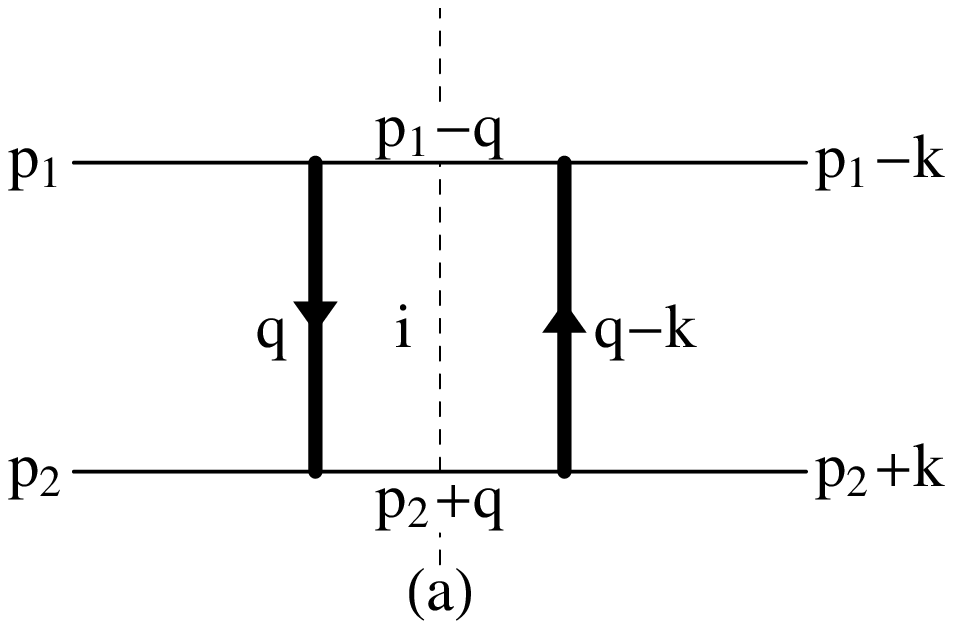,width=7cm}\hfill
      \epsfig{file=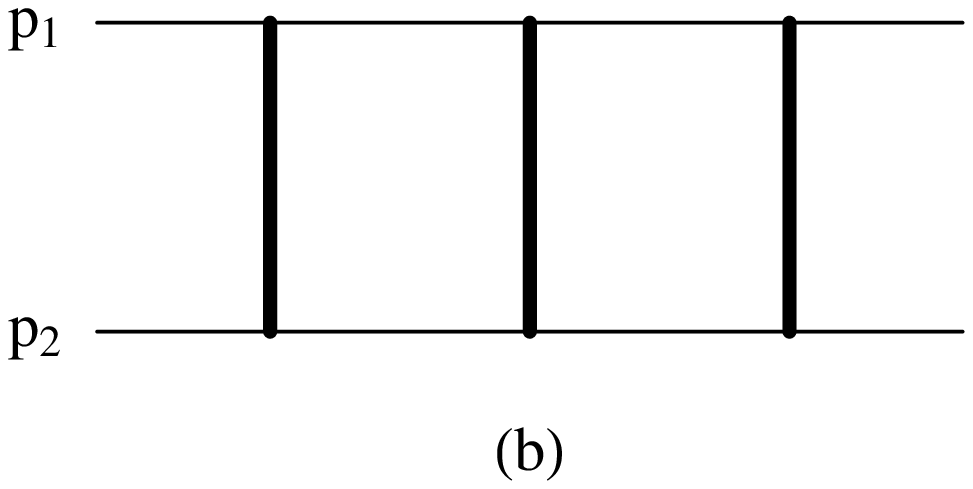,width=7cm}
\caption{\label{p6:fig:Highs} 
(a) The one loop contribution corresponding to exchange of two KK modes. 
The KK modes are drawn as thick lines and standard model particles as 
thin lines. 
(b) The two-loop contribution.
}}

For $\mpl \sim 1$ TeV the parameters of this equation are such that 
we will see a transition between the classical and 
quantum domain at LHC, and a more careful calculation, summing up 
amplitudes from ladders of t-channel exchange in \figref{p6:fig:Highs} 
to all orders is necessary. 
This calculation was performed in \cite{Giudice:2001ce} by simply ignoring the divergences 
corresponding to local contributions in \eqref{p6:eq:mint}, and recently in \cite{Sjodahl:2006gb} by 
a more careful analysis using the effective propagator \eqref{p6:eq:SMprop}. 

The parameter $b_c$ also corresponds to the impact parameter where the 
eikonal scattering phase 
\begin{eqnarray}
  \chi(b) &=& \frac{1}{2s} \int \frac{d^2 \kperp}{(2\pi)^2}\,
  e^{-i\kperp \bperp} A_{\rm{Born}}(-\kperp^2)
  \label{p6:eq:chiv1}
\end{eqnarray} 
becomes large compared to $\hbar$. This makes perfect sense, as $b_c$
represents the impact parameter separating quantum mechanical and classical
scattering. 

At least in the eikonal region, i.e. for small scattering angles, 
where there is no spin dependence, the Born amplitude can be 
written \cite{Sjodahl:2006gb}
\begin{equation}
  \Aborn(k^2=t) 
  = \frac{s^2}{2^{n-3}\pi^{n-1}\mpl^{n+2}}\, S_n \int_{0}^{\infty} 
  \frac{dm \,m^{n-1}}{k^2-m^2}\, e^{-m^2/\Ms^2}
  \label{p6:eq:ABorn}
\end{equation}
where the suppression factor $e^{-m^2/\Ms^2}$ comes from implementing the
requirement that the standard model particles live on a finite brane
\cite{Sjodahl:2006gb}. 
The same effect can be obtained by assuming a finite brane tension 
\cite{Bando:1999di}.
Computing the integrals in \eqref{p6:eq:chiv1} \cite{Sjodahl:2006gb} then 
gives the result
\begin{eqnarray}
  \label{p6:eq:chi}
  \chi(b)=-\frac{s \Ms^n}{(2\sqrt{\pi})^n \mpl^{n+2}}
  \Gamma(\frac{n}{2})
  U(\frac{n}{2},1,\frac{\Ms^2 b^2}{4})
\end{eqnarray}
where the $U$-functions are confluent hyper-geometric functions 
of the second kind.

In the limit of large third argument, $\Ms b \gg 1$ in $U$, ie.\ impact
parameters much larger than the brane width, $\chi$ can be written
\begin{eqnarray}
  \chi(b) 
  \approx 
  -\left(\frac{b_c}{b}\right)^n
  \,\,\,\,\mbox{for}\,\,\,\,
  b_c
  &\equiv&\frac{1}{\sqrt{\pi}}\left[ \frac{s \Gamma(n/2)}
    {\mpl^{n+2}}\right]^{1/n}.
  \label{p6:eq:chibigb}
\end{eqnarray}
At least if $b_c \gg 1/\Ms$ the eikonal \eqref{p6:eq:chi} reaches 1 in
the region where it is determined by \eqref{p6:eq:chibigb} and $b_c$ is
indeed the parameter associated with the transition from the quantum
mechanical to the classical region. For $\Ms$ small compared
to $\mpl$ the brane width is more important and there is an energy range 
where the impact parameter for which $|\chi|$ reaches 1, is
given by the small argument limit in $U$, rather than the large
argument limit,
\begin{eqnarray}
  \chi(b) 
  &\approx&
  \frac{2 s}{(2\sqrt{\pi})^n \mpl^2} \left(\frac{\Ms}{\mpl}\right)^n
  \left( \ln(\Ms b) +\frac{1}{2} \psi(\frac{n}{2}) \right)
  \label{p6:eq:chismallb}
\end{eqnarray}
where $\psi(\frac{n}{2})$ is the digamma function. 
The transition between,
$|\chi(b)|\approx 1$ described by \eqref{p6:eq:chismallb}, and 
$|\chi(b)|\approx 1$ described by \eqref{p6:eq:chibigb},
occurs roughly at the energy where $b_c=1/\Ms$, and the phenomenology 
will therefore differ in the regions $b_c>1/\Ms$ and $b_c<1/\Ms$.
Solving $b_c=1/\Ms$ we find 
\begin{equation}
  s_{c} = 
  \frac{\mpl^{n+2}\pi^{n/2}} {\Ms^n \Gamma(\frac{n}{2})}
  \label{p6:eq:scd}
\end{equation}
this is the line separating region 1 and 2 in \figref{p6:fig:sMs}. For 
$t \gg \Ms^2$ the all order eikonal amplitude is given by 
\begin{equation}
  \label{p6:eq:Aeik}
  A_{\rm{eik}}(k^2)=-2 i s \int d^2 \bperp e^{i \kperp\cdot \bperp}(e^{i\chi}-1).
\end{equation}
When $|\chi|$ is large compared to 1 which, for $b_c\gg1/\Ms$, happens
for $b<b_c$, the exponentiation in \eqref{p6:eq:Aeik} is important while 
for larger $b$ the eikonal amplitude is approximated by the Born term.

For $b_c<1/\Ms$, region 2 in \figref{p6:fig:sMs}, $|\chi|$ is smaller than 1
except for very small impact parameters, 
 \begin{equation}
  b < 
  \frac{1}{\Ms}
  \exp\left(- \frac{(2\sqrt{\pi})^n \mpl^{n+2}}{2 s \Ms^n}  \right),
  \label{p6:eq:bl}
\end{equation}
found by ignoring the digamma function in \eqref{p6:eq:chismallb}.
In the whole of region 1 and 2 for $t \gg \Ms^2$ the gravitational 
cross section is obtained from the all order eikonal amplitude in 
\eqref{p6:eq:Aeik} (although higher order corrections are only 
important region 1). It is given by
\begin{equation}
  \label{p6:eq:seik}
 \frac{d\sigma_{\rm{eik}}}{dt}
 =\frac{1}{16 \pi s^2} |\Aeik|^2.
\end{equation} 
If, on the other hand, $\sqrt{s} \ll \Ms$, such that $\sqrt{-t}$ 
necessarily is small compared to $\Ms$, the Born amplitude is
(apart from large angle spin dependences) fairly isotropic.
The ladder-type diagrams in \figref{p6:fig:Highs} will effectively turn 
into $\phi^4$ interactions as in \figref{p6:fig:Lows}.

\FIGURE[t]{%
      \epsfig{file=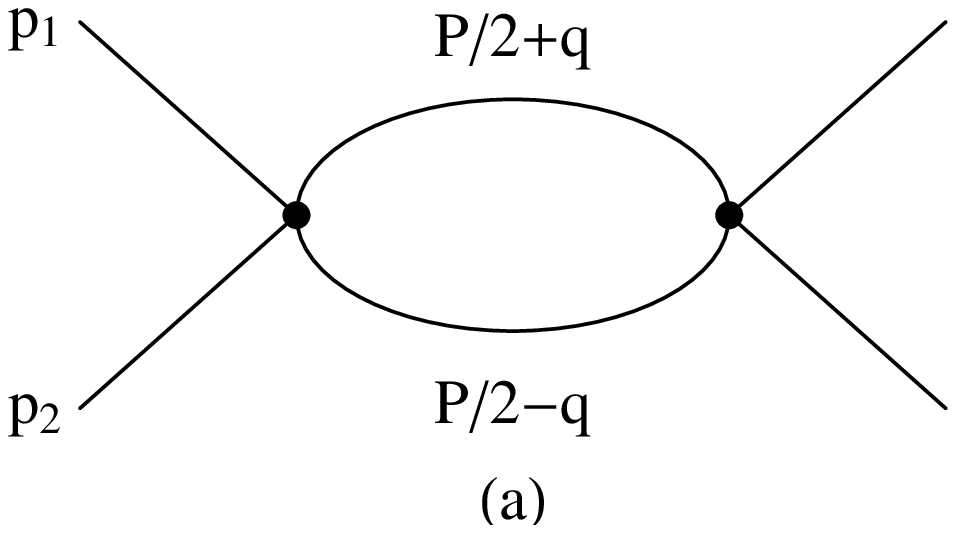,width=7cm}\hfill
      \epsfig{file=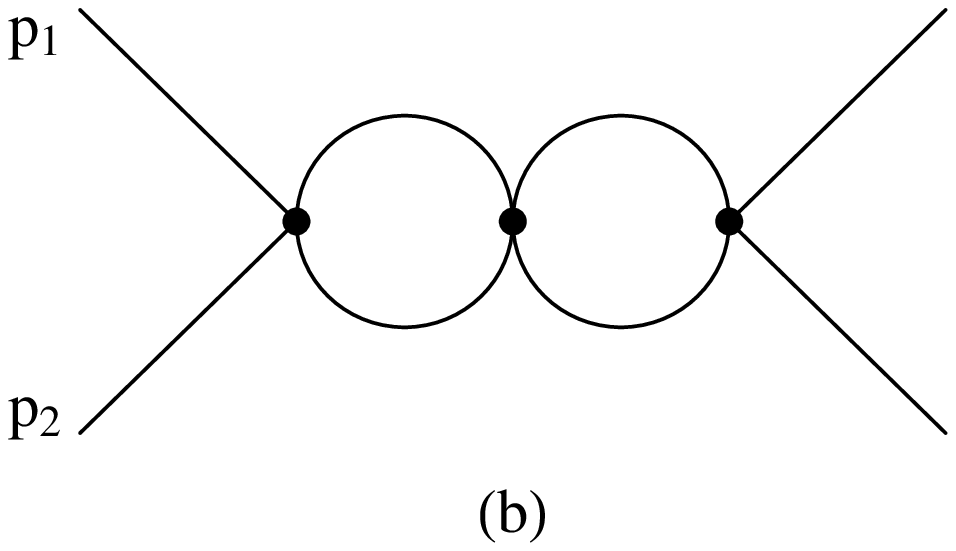,width=7cm}
\caption{\label{p6:fig:Lows} 
When the exchanged momentum is small compared to $\Ms$,
the KK propagator are effectively replaced by vertex factors.
The diagrams in fig.\ \ref{p6:fig:Highs} can then be drawn as above
with only standard model particle lines. 
}}
Since the coupling grows with energy, higher order corrections will for some 
$s$ become necessary to ensure unitarity. Summing up all contributions 
of the type in \figref{p6:fig:Lows}, neglecting large angle spin dependence, 
a geometric series is found 
\cite{Sjodahl:2006gb} which helps unitarizing the cross section. 
For the 1-loop contribution we have, with $P=p_1+p_2$ as in \figref{p6:fig:Lows},
\begin{eqnarray}
  \label{p6:eq:1-loop}
  A_{\rm{1-loop}}&\approx& \frac{-i}{2} 
  \int_{q<\Ms} \frac{d^4 q}{(2\pi)^4} A_{\rm{Born}}^2 
  \frac{1}{(P/2-q)^2} \frac{1}{(P/2+q)^2}  = \nonumber \\
  &\equiv& A_{\rm{Born}}^2\cdot X \,\,
  \mathrm{with} \,\, 
  X \approx  \frac{1}{32 \pi^2} ( \ln\frac{\Ms^2}{s/4} + i\pi) 
  \label{p6:eq:X}
\end{eqnarray}
and higher loop corrections give similar results. Summing all ladders we obtain
\begin{equation}
  A_{\mathrm{ladders} } = A_{\rm{Born}}\, (1+A_{\rm{Born}}X+
  (A_{\rm{Born} }X)^2 + \ldots\ ) 
  = \frac{ A_{\rm{Born} } }{ 1-A_{\rm{Born} }X}.
  \label{p6:eq:Xall-loop}
\end{equation}
Thus loop contributions of this type helps unitarizing the cross
section in region 4 in \figref{p6:fig:sMs}, defined to be the region
where $\sqrt{s}<\Ms$, but on-shell intermediate states in 
\figref{p6:fig:Lows} dominate.
While we can not prove that these are the most important
contributions, it seems likely as long as the cross section is
dominated by on-shell states.  In region 5 in \figref{p6:fig:sMs} this is
no longer true since here the imaginary part in \eqref{p6:eq:X} is smaller
than the real part.  The simulations performed in this paper are in the
phase space region 1, 2, 3 and 4 in \figref{p6:fig:sMs}, and we use
\eqref{p6:eq:X} to unitarize the cross section in regions 3 and 4 
(although it's not important in region 3).

\subsection{u-channel}

In the regions 1 and 2, the u-channel contribution (here generally 
defined to be the case where the outgoing particle lines are crossed 
compared to the incoming) is small compared to forward t-channel scattering. 
In the regions 3 and 4, corresponding to
$\sqrt{s}<\Ms$, it is, however, of the same order of magnitude. In fact
there is no difference between the u and t-channel in \figref{p6:fig:Lows}. 
This implies that u-channel contributions run into problems with unitarity 
at roughly the same energy as t-channel contributions, and the 
result \eqref{p6:eq:Xall-loop} (again neglecting spin dependence) can be used 
also for the u-type ladders.

The relevance of the u-channel ladders contribution is, however,
significantly lowered by the fact that interaction among identical
partons is suppressed at LHC. 
To get a handle
on the importance of u-channel contribution, assume that only valence
quarks contribute to the cross section.  This is a reasonable
assumption at sufficiently high momentum fractions and also, it will give
an upper limit. The probability for the colliding partons to have
identical flavor, spin and color is then approximately $1/10$.  

\subsection{s-channel}

For $s\ll \Ms^2$, the factor $k^2=s$ in \eqref{p6:eq:ABorn} is insignificant 
compared to most contributing KK masses and gives an amplitude similar to the 
t- and u-channels. There is, however, one complication. Due to the 
relative difference in sign between $k^2=s$ and $m^2$ in 
\eqref{p6:eq:ABorn}
KK modes can be produced on shell.
%
%

From the point of view of inclusive observables, these s-channel on-shell 
Kaluza--Klein states are, however, unimportant. 
The width of a \emph{single} KK mode with mass $m$ to decay into two 
standard model particles of energy $m/2$ is $\sim m^3 G_{N(4)}$ giving 
lifetimes of order 1000 seconds \cite{Giudice:1998ck}. These KK modes 
will leave the detectors unseen.


\subsection{Phenomenology of low energy gravitational scattering}

As already mentioned section \ref{p6:sec:div}, a gravitational scattering
where the Kaluza-Klein mode is not in the outgoing state, comes with
a momentum cut-off from the width of the brane, or from fluctuations of the
brane. In the low-energy region, 3 in \figref{p6:fig:sMs}, where
the born approximation is applicable, a cut-off dependent amplitude can
be used for describing the interaction. From the point of view of
perturbative gravitational scattering with internal KK modes only, 
this does not result in any
extra parameters to describe the interaction. Instead it suffice to
replace the Planck scale $\mpl$ by an effective Planck scale according
to
\begin{equation}
  \Meff=\frac{1}{2}
  \left( \frac{(n-2) 2^n \pi^{\frac{n-2}{2}} \mpl^{n+2}}
       {\Ms^{n-2}} \right)^\frac{1}{4}
  \label{p6:eq:Meff}
\end{equation}
such that the Born amplitude, \eqref{p6:eq:ABorn}, (neglecting spins) can
be written
\begin{equation}
  \Aborn=-\frac{s^2}{\Meff^4},
\end{equation}
after integration over $m$, neglecting $k^2=t$, $u$ or $s$.
In this kinematical region, gravitational scattering in the ADD model is a well
behaved effective field theory depending on only one free parameter, \Meff.
The low-energy spin dependent footprint of the ADD scenario for 
\textit{any} number of extra dimensions can then be written
\begin{eqnarray}
  \label{p6:eq:sigma}
  \frac{d\sigma}{dt} 
    = \frac{k_s}{s} 
    \left[ 
      \frac{\pi \alpha_s^2}{s }f(z) 
      -\frac{s\alpha_s}{\Meff^4}   g(z)
      +\frac{s^3}{\pi \Meff^8} h(z) \right]
\end{eqnarray}
where $\alpha_s$ is the strong coupling constant and
$k_s$, $g(z)$, $h(z)$ and $f(z)$ are process dependent functions taking 
spin-dependence into account given in \cite{Atwood:1999qd}.

Gravitational scattering differ from standard-model and most beyond-standard-model processes in several ways.  The experimentally most
striking is probably that it increases with increasing energy. As the
experimental situation stands today this is, however, badly
overcompensated by the decreasing parton distribution functions for
high momentum factions. The interaction is mediated by the large
number of Kaluza--Klein modes, implying that the cross section will
not have a single resonance structure, as opposed to cross section
signatures of most other beyond-standard-model particles. Due to the
different spin dependence of gravitational scattering the angular distribution
will also differ.

We will here consider another difference, namely that contrary to the
main contribution to inclusive cross sections, both in the standard
model and in super-symmetric extensions, the gravitational
interaction is colorless.  As we shall see, this implies noticeable
differences in particle multiplicity outside the jets. 

\section{Black holes in ADD scenario}
\label{p6:sec:BH}

Black holes with mass large compared to the fundamental Planck scale,
but with radius small compared to the compactification radius are expected to 
behave much like extra dimensional versions of astronomical 3+1-dimensional 
black holes. The Schwarzschild radius is given by
\begin{eqnarray}
  \label{p6:eq:rSch}
  \rsch = \frac{1}{\sqrt{\pi} \mpl}
  \left[ \frac{\MBH}{\mpl} \frac{8 \Gamma(\frac{n+3}{2})}{n+2} \right]^{\frac{1}{n+1}}
\end{eqnarray}
and the temperature is given by \cite{Myers:1986un}
\begin{eqnarray}
  \label{p6:eq:T}
  T = \frac{n+1}{4 \pi \rsch}.
\end{eqnarray}
Note that small black holes are hotter.

A major difference between black holes in the ADD scenario and
ordinary 3-dimensional black holes is that ADD black holes do not
radiate gauge fields into most of phase space, since only gravity is
allowed to propagate in the extra dimensions. 
One may believe that this would lead to
almost no radiation on the brane (where gauge fields and, hence, also we
live) as the bulk phase space is much larger. However, it has been
shown that this is not necessarily the case
\cite{Emparan:2000rs}.

Since we are considering the non-idealized situation of a finite brane
width we must also consider the implications of this on black hole
production. In particular, a natural requirement is that the brane is
not more extended than the black hole, leading to the condition $\rsch
< 1/M_b$ for the formation of black holes. As we will see this prevents 
black holes from appearing at the LHC for sufficiently small \Ms.

On the other hand, if \Ms\ is large, we may, with increasing $\sqrt{s}$,
go directly from the Born region 3 in \figref{p6:fig:sMs} to black hole 
production. This should be worrying since the black holes are 
treated semi-classically but the gravitational scattering in region 3
is purely quantum mechanical. It is reasonable that the 
black holes should start behaving classically 
when the Compton wave length is of the same order as the black hole radius,
but it would have been more comforting to only study black hole production
in region 1 in \figref{p6:fig:sMs}, where the gravitational scattering is 
mainly classical already at lower energies.
This represents a genuine quantum 
gravity uncertainty.

Already at a classical level the cross section for black hole creation is
subject to significant uncertainties. This is basically due the fact that 
it does not suffice to consider the colliding objects, but in addition the  
curvature of space-time far outside the black hole need to be calculated. 
Classical numerical simulations for black hole formation in extra dimensions 
have been performed in \cite{Yoshino:2002tx,Yoshino:2005hi} with the result 
that the geometric cross section, $\pi r^2$, should be multiplied with a 
factor $\sim 0.7-3$, increasing with the number of extra dimensions.
For this paper we have, however, chosen to keep the constant at 1.

As the black holes considered here are formed from partons inside the 
protons there is also an uncertainty from the usage of parton distribution
functions for an essentially non-perturbative process \cite{Harris:2004xt} .
(A discussion about the effects of quantum fluctuations 
based on wave packages can be found in \cite{Giddings:2004xy, Rychkov:2004sn}.) 

Then there is the question of the onset of black hole production. It can be
argued that no black holes should be formed below (roughly) the Planck scale 
as the uncertainty principle would forbid sufficient localization of the 
partons. But precisely when does black holes begin to form? 

We consider first the condition that the black holes have to be 
well localized in our ordinary dimension.
Looking at the momenta of the incoming partons in their combined rest 
frame it is reasonable to  require that their
wavelength, $\lambda_l\propto 2/\sqrt{s}$, is less than
\rsch. The corresponding requirement in the transverse direction gives
the requirement: $\lambda_\perp \propto 1/ p_T< \rsch$. Clearly one can argue 
about the proportionality constant. 
We have chosen
\begin{eqnarray}
  \label{p6:eq:Mmin1}
  \mmin=2/\rsch(\mmin).
\end{eqnarray}
Combining this with the expression for the Schwarzschild radius 
\eqref{p6:eq:rSch} we get 
\begin{eqnarray}
  \label{p6:eq:ratio2}
  \mmin={\mpl}
           \left[ \frac{(2\sqrt{\pi})^{n+1}(n+2)}
             {8 \Gamma(\frac{n+3}{2})} \right]^{\frac{1}{n+2}}.
\end{eqnarray} 
Numerically the value of \mmin\ is then approximately twice the Planck mass.

As we consider a finite brane width we must add the condition
$\rsch>1/M_s$, leading to the minimal mass
\begin{eqnarray}
  \label{p6:eq:Mmin2}
  \mminn{2}=\frac{\mpl^{n+2} (2+n) \pi^{\frac{n+1}{2}}}
       {8 \Gamma \left[ \frac{3+n}{2} \right] M_s^{1+n}}.          
\end{eqnarray}
Again one can argue about the proportionality constant. 
While we take into account the effects of a finite brane, we do
not consider the dynamics governing the brane, and possibly describing 
its width, although this may have significant effects on the spectra 
observed \cite{Frolov:2004wy, Frolov:2004bq}.

Once a black hole has formed it is believed to lose most of its geometric
asymmetries in a short period referred to as the balding phase. 
This phase leaves a black hole whose only geometric asymmetry can be 
described by one angular momentum parameter. However, it turns out that this 
angular momentum tends to be lost rather quickly via Hawking radiation, such 
that the black hole (apart from gauge charges) can be described by the 
Schwarzschild metric.

Neglecting the gauge charges, which in the case of electromagnetism has
been shown to have a modest influence \cite{Page:1977um}, the disappearance of 
the black hole would be well described by Hawking radiation if the 
black hole was much heavier than the Planck mass, and if no brane effects,
such as the black hole recoiling of the brane 
\cite{Frolov:2002gf,Frolov:2002as}, or interacting with the brane 
\cite{Frolov:2004wy,Frolov:2004bq} is taken into account. The problem is that
most collider-produced black holes will not be much heavier than the Planck mass.

For a hole which is not heavy compared to the Planck mass one cannot 
treat the metric as a static background for the emitted quanta, the back-reaction of the quanta to the metric should be taken into account and this
is not done in the derivation of the Hawking radiation \cite{Hawking:1974sw}.
Also, at some point, the lifetime of the black hole becomes shorter
than its radius. This makes it difficult to talk about a thermalized black 
hole.

Considering all of this, it should not come as a surprise if black holes 
where observed with spectra which differs significantly from that 
expected from \eqref{p6:eq:T}.

\section{Results}
\label{p6:sec:results}

We have used the amplitudes for gravitational scattering for the
different regions in \figref{p6:fig:sMs} presented above to reweight the 
standard QCD $2\to2$
scatterings in the \pythia (version 6.2 \cite{Sjostrand:2001yu}) event
generator. 
In the regions 1 and 2 we have used the (elastic spin-independent t-type) 
all order eikonal cross 
section from \eqref{p6:eq:chiv1} and \eqref{p6:eq:Aeik}. 
In the regions 3 and 4 we have used 
the spin dependent Born amplitude \cite{Giudice:2004mg} 
corresponding to \eqref{p6:eq:sigma}, 
and higher order corrections according to \eqref{p6:eq:Xall-loop}.
In region 3 the higher order corrections are small, but in
region 4 they are essential. In the case of particle-antiparticle
scattering, such that the scattering can be mediated via the s-channel,
we have ``unitarized'' also the s-channel contribution in region 4 (and 3)
using \eqref{p6:eq:Xall-loop}, despite that fact the the s-type ladders,
diagrams in \figref{p6:fig:Lows} rotated by $\pi/2$ do not have on-shell 
intermediate standard model particles. 
There are thus several fundamental uncertainties 
associated with gravitational scattering in region 4. First, we
use the spin dependent Born amplitude, but we do not take spin dependence 
consistently into account in \eqref{p6:eq:Xall-loop} since we use the 
same $\Aborn(t)$ everywhere in all ladders. 
Second, we suppress s-type contributions in the same way as u- and t-type.
(Note that we call the ladders in \figref{p6:fig:Highs} and \figref{p6:fig:Lows}
t-type, sometimes these diagrams are referred to as s-channel, since the 
resummation is in $s$.) 

The different treatments in the various regions means that we could expect 
a discontinuous
transition when $\sqrt{s}$ is increased, such that we cross the line
$\sqrt{s}=M_s$ in \figref{p6:fig:sMs}. As long as the Born approximation
is applicable (regions 2 and 3), this transition just corresponds
to starting neglecting spin-dependence in region 2. If the transition
is between region 4 and 1, the situation is, however, worse due to
fundamental uncertainties associated with region 4.

For each generated $2\to2$ scattering we also change the
color flow between the scattered partons with a probability
$\sigma\sub{ADD}/(\sigma\sub{ADD}+\sigma\sub{QCD})$ to reflect the
colorless nature of the graviton exchange. The resulting partonic
state is then allowed to evolve a QCD cascade and is finally
hadronized to produce fully simulated hadron-level events. Where relevant, we have
also added multiple soft and semi-hard QCD scatterings to simulate the
underlying event according to the model implemented in
\pythia \cite{Sjostrand:1987su}. 

In addition, we have used the \charybdis\cite{Harris:2003db} program to 
simulate the production and decay of black holes as described in section \ref{p6:sec:BH} and in \cite{Lonnblad:2005ah}. 
To ensure that the energy is sufficiently localized, in our ordinary dimensions
and in the extra dimensions, we have required a minimal black hole
mass according to \eqref{p6:eq:ratio2} and \eqref{p6:eq:Mmin2}.
We also use the Schwarzschild radius to
cut off any QCD and gravitational $2\to2$ scatterings in region 3 and 4
for large enough masses and transverse momenta as discussed in 
\cite{Lonnblad:2005ah}. In region 1 (and 2) we use the impact parameter description 
defined via \eqref{p6:eq:chi} to turn off gravitational interactions at 
distances smaller than \rsch. Clearly this simpleminded approach of turning 
of gravitational scattering should not be seen as the final word. In 
particular our understanding of gravitational scattering, and hence its 
turnoff, is limited in region 4.

We limit our investigation to two standard inclusive high-\ET\ jet
observables \cite{Lonnblad:2005ah,Harris:2004xt,Humanic:2006xg}, namely the \ET-spectrum of the highest-\ET\ jet in an
event, and the distribution in invariant mass, \Mjj, of the two highest-\ET\ 
jets in an event. As high-\ET\ jets will be a part of almost any
signal of new physics at the LHC, such observables will be measured
early on after the start of the experiments and it is also where one
would expect gravitational scatterings to contribute. We use a simple
cone algorithm\footnote{The \texttt{GETJET} algorithm originally
  written by Frank Paige.} with a code radius of 0.7, assuming a
calorimeter covering the pseudo-rapidity interval, $|\eta|<2.5$, and
requiring a minimum \ET\ of $100$~GeV for the resulting jets. 
We have checked that our results do not depend much on the algorithm
chosen.

\TABLE{
  \begin{tabular}{|l|c|c|r|r|}
    \hline
    \Meff & $n$ & $\Ms/\mpl$ & \mpl & \Ms\\
    \hline
    1.0 & 4 & $\frac{1}{2}$ & 0.45 & 0.22\\
    1.0 & 4 & 1 & 0.63 & 0.63\\
    1.0 & 4 & 2 & 0.89 & 1.79\\
    1.0 & 4 & 4 & 1.26 & 5.05\\
    1.0 & 6 & 2 & 0.56 & 1.13\\
    0.7 & 4 & 4 & 0.88 & 3.54\\
    4.0 & 4 & 4 & 5.05 & 20.21\\
    \hline
  \end{tabular}
  \caption{\label{p6:tab:mpms} The different values of \Meff, number
    of extra dimensions, $n$, and the ratio of $\Ms/\mpl$ used in the
    simulations together with the resulting approximative values of
    \mpl\ and \Ms. The masses are all given in units of TeV.}}

\FIGURE[t]{
  \epsfig{file=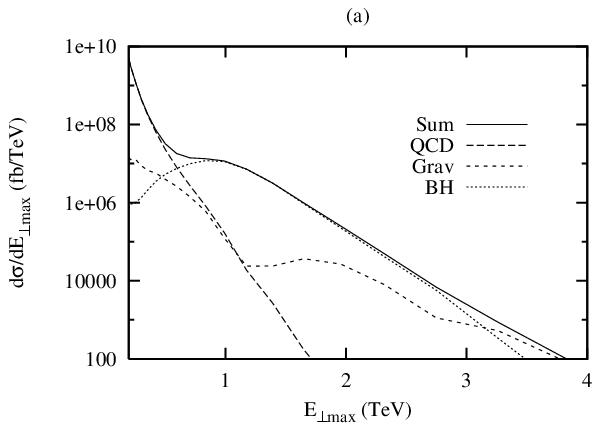,width=7.4cm}
  \epsfig{file=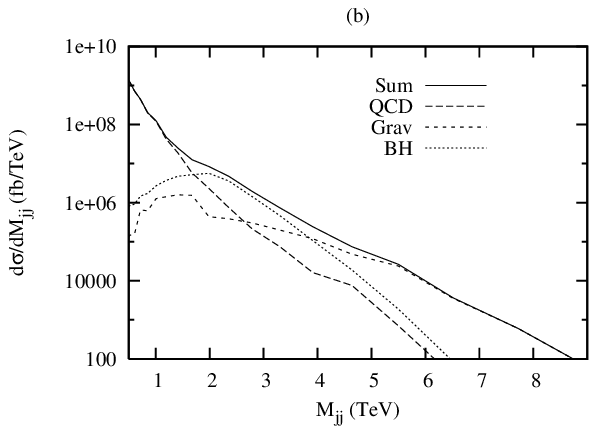,width=7.4cm}
  \caption{\label{p6:fig:ET1} (a) The \ET-spectrum of the highest
    \ET\ jet in an event, and (b) the invariant mass spectrum of
    the two highest \ET\ jets in an event at the LHC. In both
    cases $\Meff=1$~TeV with 4 extra dimensions and
    $\Ms/\mpl=2$ ($\mpl\approx0.9$~TeV, $\Ms\approx1.8$~TeV).
    The long-dashed lines are the contribution from QCD scatterings,
    short-dashed lines the contribution from gravitational
    scatterings, dotted lines the contribution from the decay of black
    holes and full lines the sum of all contributions.}  }

In \figref{p6:fig:ET1} we show generated the \ETI{max}\ and \Mjj\
distributions at the LHC for the case of four extra dimensions,
$\Meff=1$~TeV and $\Ms/\mpl=2$ (see table \ref{p6:tab:mpms} for the
resulting values of \mpl\ and \Ms). In the \ETI{max} spectra we see
that the cross section is dominated by QCD scatterings at low \ET as
expected, followed by an intermediate region where gravitational
scattering becomes important before black-hole production starts
dominating the cross section at large \ET. For the \Mjj-spectrum,
the situation is different, and the gravitational scatterings
dominates at large masses.

Modulo effects of the parton densities we expect both gravitational
scattering and black-hole production to increase with energy.  
For black holes one may naively not necessarily expect to find 
high-\ET\ jets, as energetic quanta are Boltzmann suppressed in the 
Hawking radiation. However, it turns out that the large cross
section for a black hole to form at high $s$, multiplied with the small
probability for the Black hole to radiate extremely energetic quanta,
may dominate over the non-black hole cross section for rather large
transverse momenta \cite{Lonnblad:2005ah}. 
(Even if well localized QCD and gravitational
scattering events are not suppressed due to black hole production.)
These extremely energetic quanta do, however, not obey the semiclassical
approximation in the Hawking radiation derivation, and are therefore associated
with large uncertainties.

For large \ET, however, the gravitational scattering events, just as
the QCD ones, may be localized inside the Schwarzschild radius and
will collapse into a black hole.

\FIGURE[t]{
  \epsfig{file=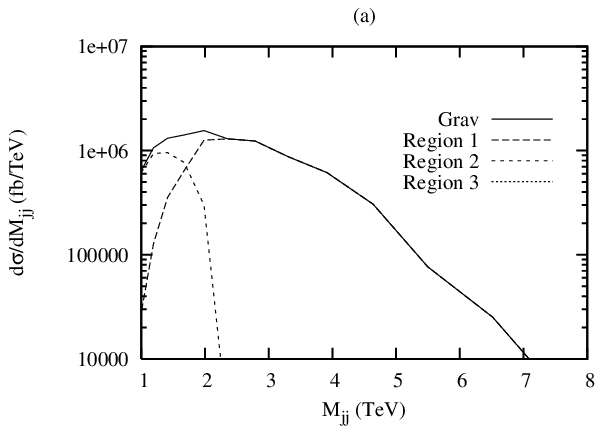,width=7.4cm}
  \epsfig{file=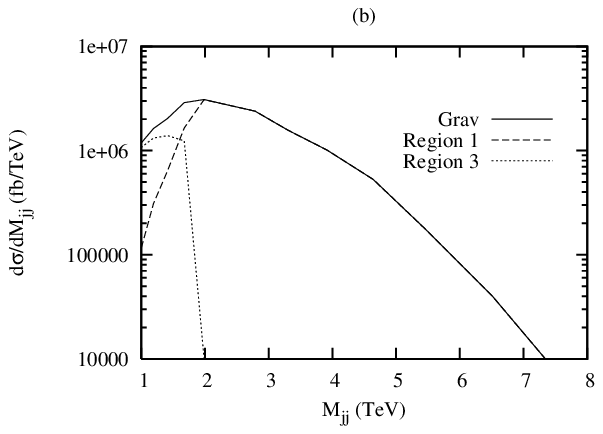,width=7.4cm}
  \epsfig{file=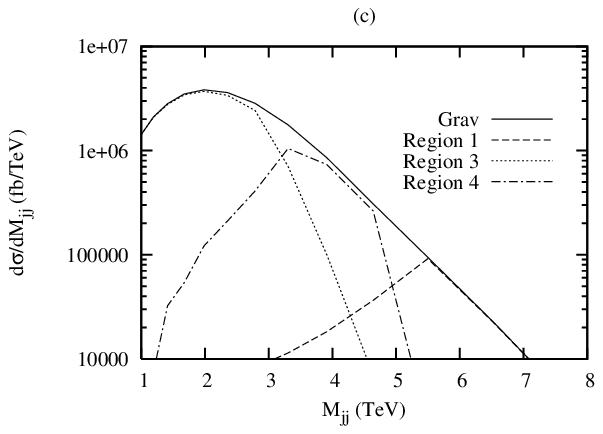,width=7.4cm}
  \caption{\label{p6:fig:MJ2} The contribution of different regions in
    figure \ref{p6:fig:sMs} to the di-jet mass spectrum from
    gravitational scatterings at the LHC with $\Meff=1$~TeV, 4
    extra dimensions and $\Ms/\mpl=1$~(a), $2$~(b) and $4$~(c). In
    all cases the full line is the sum of all contributions and the
    contributions from regions 1, 2, 3 and 4 is given by the
    long-dashed, short-dashed, dotted and dash-dotted lines
    respectively. }}

In \figref{p6:fig:MJ2} we show the \Mjj-distribution of gravitational
scatterings only, divided into the contributions from the different
regions in \figref{p6:fig:sMs}. Keeping $\Meff=1$~TeV and the number of
extra dimensions (4) fixed, we vary $\Ms/\mpl$ and find that the
contribution from region 1 dominates except in the low-mass regions
below \Ms. The transitions between the regions are not sharp, mainly
due to the smearing introduced by shower, hadronization and the jet
reconstruction. 
This smearing hides the fact that the transition between  
$\sqrt{s} > \Ms$ and $\sqrt{s} < \Ms$ is discontinuous in the distribution
of the generated $s$. In the case this transition occurs between
region 2 and 3, where the Born approximation is applicable, the 
discontinuity is not even visible in the generated $s$-distribution. If the 
transition occurs between region 4 and 1, a discontinuity can, however, 
be seen.

\FIGURE[t]{
  \epsfig{file=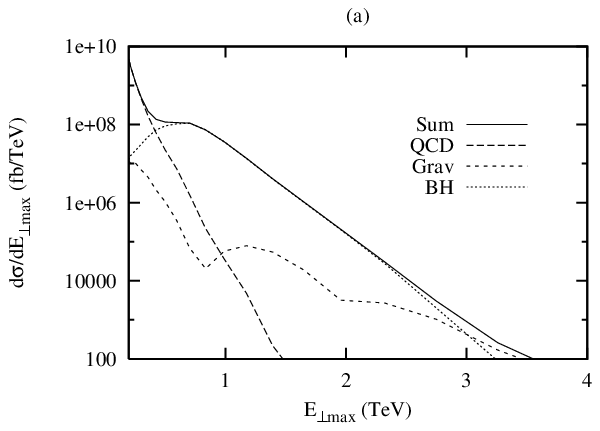,width=7.4cm}
  \epsfig{file=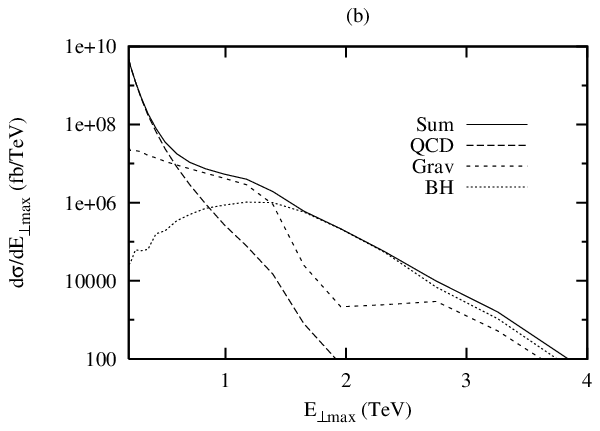,width=7.4cm}

  \epsfig{file=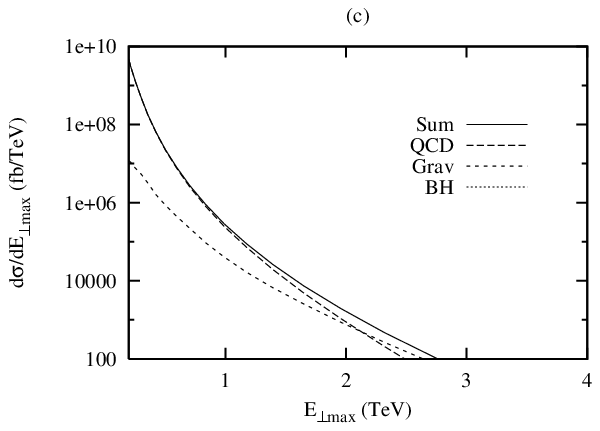,width=7.4cm}

\caption{\label{p6:fig:ET3} The same as \figrefi{p6:fig:ET1}a, but with
  $\Ms/\mpl=1$~(a), $4$~(b) and $0.5$~(c).}}

We note in table \ref{p6:tab:mpms} that, although \Meff\ is kept fixed,
giving the same amount of gravitational scattering at $\sqrt{s}\ll M_s$,
increasing the ratio $\Ms/\mpl$ will increase \textit{both} \Ms\ and \mpl.
And since black-hole production depends of \mpl\ and $M_s$ differently via
\eqsref{p6:eq:rSch}, (\ref{p6:eq:Mmin1}) and (\ref{p6:eq:Mmin2}), 
we can vary the relative importance of gravitational scattering and 
black-hole production by varying $\Ms/\mpl$. Hence we see in 
\figrefi{p6:fig:ET3}a that lowering $\Ms/\mpl$
to 1, the gravitational scattering will never give a sizeable
contribution to the \ETI{max}-distribution, while increasing the ratio
to 4 (\figrefi{p6:fig:ET3}b) results in the gravitational scattering
dominating the cross section further out in \ET\ as compared to
\figrefi{p6:fig:ET1}a. In \figrefi{p6:fig:ET3}c we decrease the ratio even
further to 0.5 which results in a brane thickness so large that black
holes can never be formed at the LHC, and the only indication of
the presence of extra dimensions in the \ET\-spectra is a slight increase in 
the cross section for large \ETI{max}.

\FIGURE[t]{
  \epsfig{file=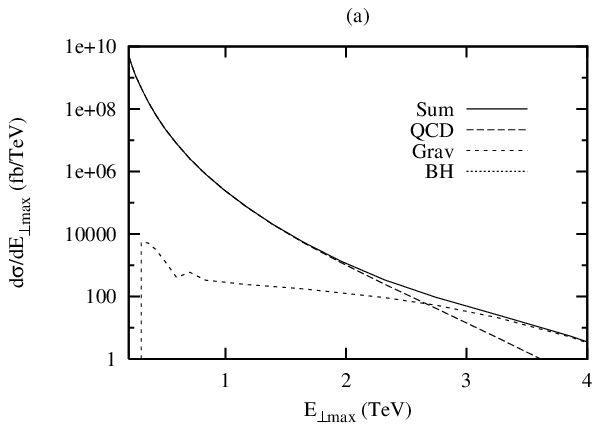,width=7.4cm}
  \epsfig{file=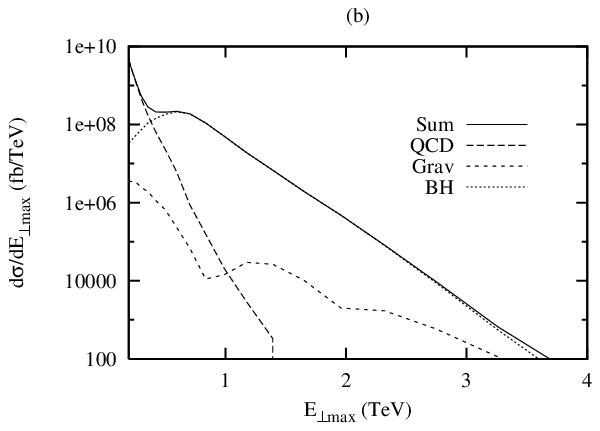,width=7.4cm}
\caption{\label{p6:fig:ET3e} The same as \figrefi{p6:fig:ET1}a, but with
  (a) 4 extra dimensions, $\Meff=4$~TeV, $\Ms/\mpl=4$ and (b) 6 extra
  dimensions, $\Meff=1$~TeV, $\Ms/\mpl=2$.}}

A similar effect can be obtained by increasing the effective mass, 
while keeping the ratio $M_s/M_p$ fixed, hence increasing 
both $\mpl$ and $M_s$. 
This is done in \figrefi{p6:fig:ET3e}a and, again,
the only visible effect of the extra dimensions is from gravitational scattering in the high-\ET\ 
region. On the other hand we see in \figrefi{p6:fig:ET3e}b how increasing
the number of extra dimensions to 6, keeping $\Meff=1$~TeV, gives a
negligible contribution from gravitational scattering to the
\ETI{max}-distribution, which instead is completely dominated by the decay of
black holes.

If large extra dimensions exist, one would hope that the scales are
such that they would be easily discovered at the LHC by, eg.\ the
striking signature of a decaying black hole. However, it is easy to
see how nature could conspire, such that the only signal in the \ET\ 
spectrum would be a
slight increase of the high-\ET\ jet cross section. There are, of
course, other signals, such as the production of real gravitons,
showing up as large missing transverse momenta. But such signals could
also be the result of other possible beyond-the-standard-model
scenarios. In any case, it would be desirable to be able to
distinguish gravitational scatterings from standard QCD events. One
obvious difference is that the exchange of a graviton is colorless in
contrast to a QCD scattering. This will necessarily give rise to a
different color topology in gravitational events as compared to QCD
ones. In particular one would expect the appearance of so-called
rapidity gaps between the jets in gravitational scattering events.
Although these gaps may be filled by secondary soft and semi-hard QCD
scatterings, one may still expect a lower activity between the jets in
such events.

\FIGURE[t]{
  \epsfig{file=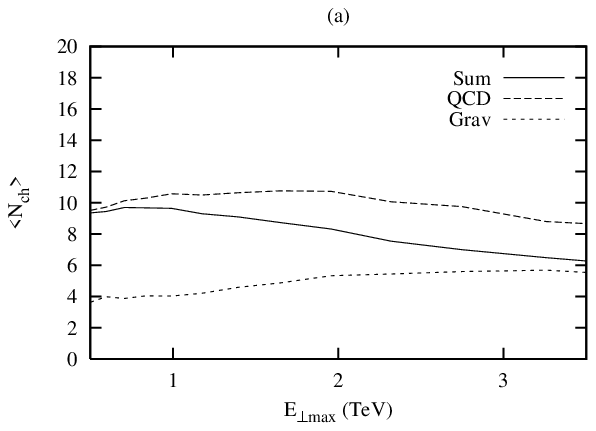,width=7.4cm}
  \epsfig{file=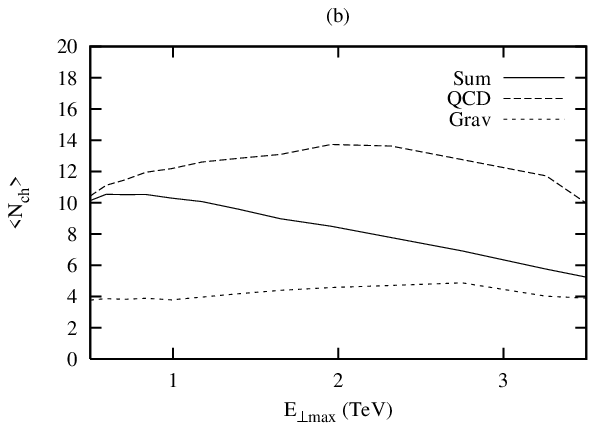,width=7.4cm}
  \caption{\label{p6:fig:ET4} The average number of charged particles
    outside the jet-cones in the central rapidity unit between the two
    hardest jets in events corresponding to \figrefi{p6:fig:ET3}c. The
    full line is for all events while the long-dashed and short-dashed
    are for QCD and gravitational scatterings respectively. In (b)
    only events with a minimum pseudo-rapidity difference of one unit
    between the two highest \ET\ jets are included, while in (a) there
    is no such requirement.}}

In \figrefi{p6:fig:ET4}a we show the average number of charged
particles with a transverse momentum above $0.5$~GeV outside the jet
cones in the middle unit of pseudo-rapidity between the two hardest
jets as a function of \ETI{max}. Hence, we count only charged particles,
$c$, with
\begin{eqnarray}
  \pTi{c}&>&0.5\mbox{~GeV},\nonumber\\
  \Delta R_{c1},\Delta R_{c2}&>&0.7\mbox{~and}\nonumber\\
  \left|\eta_c-\frac{\eta_1+\eta_2}{2}\right|&<&0.5,
  \label{p6:eq:midrap}
\end{eqnarray}
where $\eta_c$ and $\eta_i$ are the pseudo rapidities of the particle
and (the center of) jet $i$ respectively and $\Delta R_{ci}$ is the
distance between the particle and jet $i$ in the
pseudo-rapidity--azimuth-angle $(\eta,\phi)$ plane.  In this
simulation we have included multiple interactions in \pythia to
simulate the underlying event.\footnote{Using parameter settings
  according to the so-called Tune-A by Rick Field\cite{RickTuneA}.} We
see that the expectation from QCD events is around 10 particles, while
for gravitational scatterings the average is around 5. With sufficient
statistics it could therefore be possible to observe the decrease in
the number of charge particles with increasing \ETI{max} as
gravitational scatterings starts to dominate. In
\figrefi{p6:fig:ET4}a we have not required a large rapidity separation
between the jets. Doing so would increase the effect, as shown in
\figrefi{p6:fig:ET4}b, but on the other hand the statistics would
decrease.

We note that the absolute numbers in \figref{p6:fig:ET4} is very
sensitive to the modeling of the underlying event, which is very
difficult to predict for the LHC. The underlying event should,
however, give the same contribution to both scattering types, and the
difference between the two should be fairly well predicted by \pythia.

\FIGURE[t]{
  \epsfig{file=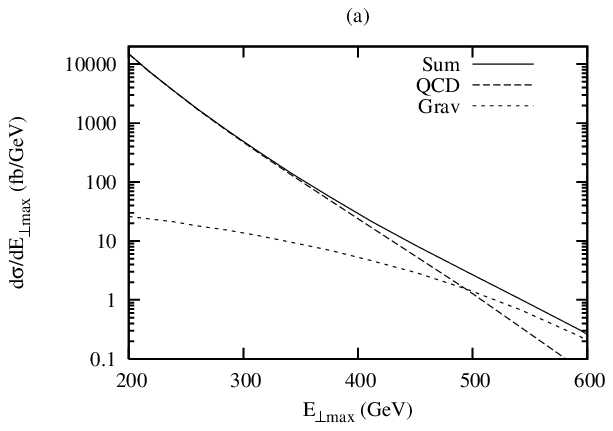,width=7.4cm}
  \epsfig{file=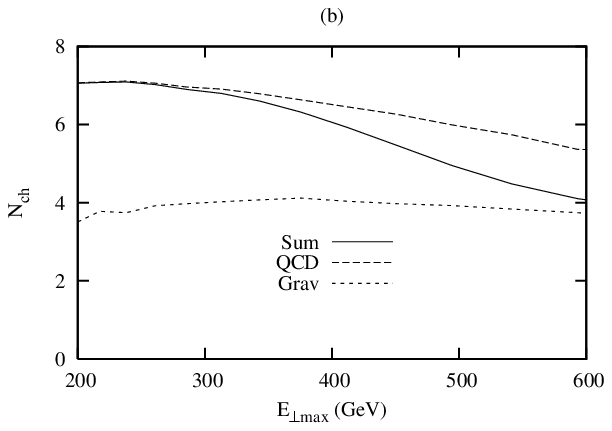,width=7.4cm}
  \caption{\label{p6:fig:ET5} (a) The \ET-spectrum of the highest
    \ET\ jet in an event, and (b) The average number of charged
    particles outside the jet-cones in the central rapidity unit
    between the two hardest jets at the Tevatron for
    $\Meff=700$~GeV, $\Ms/\mpl=4$ and 4 extra dimensions. The
    full line is for all event while the long-dashed and short-dashed
    are for QCD and gravitational scatterings respectively.}}

At the Tevatron there was an indication of an excess of the cross
section for very high \ET\ jets as compared to the QCD prediction
\cite{Abe:1996wy,Affolder:2001fa}. Although a re-evaluation of the
uncertainties due to the parton density parameterizations used in the
QCD calculations has brought this excess within the limits of the
statistical and systematical errors, it is still intriguing that such
an excess could be the signal of the onset of gravitational scattering
due to the presence of large extra dimensions.

In \figrefi{p6:fig:ET5}a we show our prediction for the \ETI{max}
distribution for $n=4$, $\Meff=700$~GeV and $\Ms/\mpl=4$ at the
Tevatron. The parameters were chosen so that the excess above standard
QCD production is approximately within the statistical and
systematical uncertainties of the corresponding Tevatron measurement.
In \figrefi{p6:fig:ET5}b we then show the average number of charged
particles outside the cones (same as in \figref{p6:fig:ET4}, but counting
charged particles with transverse momenta down to $0.25$~GeV). The
decrease in the region where gravitational scatterings become
important is significant, although it may be difficult to get enough
statistics to measure it even for Run-II at the Tevatron. However, it
is not completely inconceivable that by finding a more sensitive
observable of the color structure, we could be able to see the first
indication of large extra dimensions already at the Tevatron, before
LHC is switched on.

\section{Conclusions}
\label{p6:sec:conclusions}

We have studied gravitational scattering and black hole production at
the LHC and the Tevatron in the ADD scenario assuming that brane on
which the standard model fields live have a finite width. We found
that the relative importance of gravitational scattering and black
hole production is sensitive to this width, and that for large widths
the extension of the standard model particle fields into the bulk, may
prevent black holes from forming since the energy may not be
sufficiently localized within the black hole radius.

A wide brane corresponds to a low cut-off for virtual Kaluza--Klein
modes (the brane width and the KK cut-off, \Ms, are inversely related
via a Fourier transform \cite{Sjodahl:2006gb}) and therefore results
in weaker gravitational interaction in the non-classical regions.
It is thus possible for
nature to conspire, by choosing a low $M_s$, such that neither much
gravitational scattering or black holes is observed at the LHC.  In
this case processes involving the production of on-shell KK modes
resulting in missing $E_{\perp}$ may become important observables.

In our simulations we have used values of \mpl, \Ms\ and the number of
extra dimensions, $n$, which we believe have not yet been excluded by
experiments (see eg.\ \cite{PDBook,Abulencia:2006kk} for recent reviews). 
Most of
these limits are only relevant to \mpl\, but restrictions on \Ms\ 
could be obtained by considering processes involving both virtual and
real KK modes.

In the case of low \Ms, only a weak increase in the jet spectra could
be observed at LHC, and the signal of missing $E_{\perp}$ could be the result 
of SUSY. We point out that the colorless nature of gravitational 
scattering could be a way of distinguishing gravity induced events from 
other beyond-standard-model extensions. In fact this method could be used to 
indicate if an excess of jet activity at high transverse energies at the
Tevatron is a result of gravitational scattering.


\appendix

\section{Appendix}
\label{p6:sec:Appendix}

There are at least four definitions of the Planck mass. Often one have
to understand which definition an author uses by the relation of the
Planck mass to the 4-dimensional Newton's constant $G_{N(4)}$ or to
the Schwarzschild radius of a black hole.  The process of hunting down
constants is further complicated by the use of different definitions
of the compactification radius, many authors \cite{Han:1998sg} mean by
the compactification radius rather the compactification circumference,
here denoted $L$, whereas others really mean the radius,
$R=L/(2 \pi)$.  In order to simplify comparison between the different
conventions we here state the relations between the Planck masses
$\mpl$ (used here) and in \cite{Harris:2003db}, $\MD$ used in
\cite{Giudice:1998ck, Giudice:2001ce}, $M_G$, $M_S$ used in
\cite{Han:1998sg} and
the 4-dimensional Newton's constant $G_{N(4)}$, the relation between
the Planck masses and Schwarzschild radius, and the relations of the
Planck masses to each other.
   
\begin{eqnarray}
  \label{p6:eq:MDGN}
  \MD^{2+n}=\frac{1}{8\pi R^n G_{N(4)}}
\end{eqnarray}

\begin{eqnarray}
  \label{p6:eq:MpGN}
  \mpl^{2+n}=\frac{1}{L^n G_{N(4)}}
\end{eqnarray}

\begin{eqnarray}
  \label{p6:eq:MGGN}
  M_G^{2+n}=\frac{2^{n-2}\pi^{n-1}}
  {L^n G_{N(4)}}
\end{eqnarray}

\begin{eqnarray}
  \label{p6:eq:MsGN}
  M_S^{2+n}=\frac{\Gamma \left(\frac{n}{2} \right) \pi^{n/2}}
  {2^{1-n} L^n G_{N(4)}}
\end{eqnarray}

\begin{eqnarray}
  \label{p6:eq:rMD}
  \rsch=\frac{1}{\MD}\left[\frac{M_{BH}}{\MD}\right]^{\frac{1}{n+1}}
  \left[
  \frac{2^n \pi^{\frac{n-3}{2}} \Gamma\left(\frac{n+3}{2} \right)}{n+2}
  \right]^{\frac{1}{n+1}}
\end{eqnarray}

\begin{eqnarray}
  \label{p6:eq:rMp}
  \rsch=\frac{1}{\sqrt{\pi}\mpl}
  \left[\frac{M_{BH}}{\mpl}\right]^{\frac{1}{n+1}}
  \left[
    \frac {8 \Gamma\left(\frac{n+3}{2} \right)}{n+2}
  \right]^{\frac{1}{n+1}}
\end{eqnarray}

\begin{eqnarray}
  \label{p6:eq:rMG}
  \rsch=\frac{2}{M_G}
  \left[\frac{M_{BH}}{M_G}\right]^{\frac{1}{n+1}}
  \left[
    \frac {\pi^{\frac{n-3}{2}} \Gamma\left(\frac{n+3}{2} \right)}{n+2}
  \right]^{\frac{1}{n+1}}
\end{eqnarray}

\begin{eqnarray}
  \label{p6:eq:rMs}
  \rsch=\frac{1}{M_S}
  \left[\frac{M_{BH}}{M_S}\right]^{\frac{1}{n+1}}
  \left[
    \frac{2^{2+n} \Gamma\left(\frac{n+3}{2}\right) \Gamma\left(\frac{n}{2} \right)}
{\sqrt{\pi}(n+2)}
  \right]^{\frac{1}{n+1}}
\end{eqnarray}

\begin{eqnarray}
  \label{p6:eq:MpMp}
  \mpl
  &=&2^{\frac{3-n}{2+n}} \pi^{\frac{1-n}{2+n}} \MD\nonumber\\
  &=&2^{\frac{2-n}{2+n}} \pi^{\frac{1-n}{2+n}} M_G\nonumber\\
  &=&2^{\frac{1-n}{2+n}} \pi^{\frac{-n}{4+2n}} 
 \Gamma\left(\frac{n}{2}\right)^{\frac{-1}{2+n}} M_S
\end{eqnarray}

\bibliographystyle{utcaps}  
\bibliography{references,refs}

\end{document}